\documentclass{nime-alternate} 

\usepackage{anonymize} 		   

\usepackage[utf8]{inputenc}

\begin{document}

\conferenceinfo{NIME'20,}{July 21-25, 2020, Royal Birmingham Conservatoire, ~~~~~~~~~~~~ Birmingham City University, Birmingham, United Kingdom.}

\title{Nuanced and Interrelated Mediations and Exigencies (NIME): Addressing the Prevailing Political and Epistemological Crises}

%
%
%
\label{key}
%

\numberofauthors{2} 
%
\author{
%
%
\alignauthor
\anonymize{Lauren Hayes}\\
       \affaddr{\anonymize{Arts, Media + Engineering}}\\
       \affaddr{\anonymize{Arizona State University}}\\
       \affaddr{\anonymize{Tempe, Arizona 85287}}\\
       \email{\anonymize{lauren.s.hayes@asu.edu}}
\alignauthor
\anonymize{Adnan Marquez-Borbon}\\
       \affaddr{\anonymize{Facultad de Artes}}\\
       \affaddr{\anonymize{Universidad Aut{\'o}noma de Baja California}}\\
       \affaddr{\anonymize{Ensenada, Mexico}}\\
       \email{\anonymize{adnan.marquez@uabc.edu.mx}}
}


\date{30 July 1999}


\maketitle
\begin{abstract}
Nearly two decades after its inception as a workshop at the ACM Conference on Human Factors in Computing Systems, NIME exists as an established international conference significantly distinct from its precursor. While this origin story is often noted, the implications of NIME's history as emerging from a field predominantly dealing with human-computer interaction have rarely been discussed. In this paper we highlight many of the recent---and some not so recent---challenges that have been brought upon the NIME community as it attempts to maintain and expand its identity as a platform for multidisciplinary research into HCI, interface design, and electronic and computer music. We discuss the relationship between the market demands of the neoliberal university---which have underpinned academia's drive for innovation---and the quantification and economisation of research performance which have facilitated certain disciplinary and social frictions to emerge within NIME-related research and practice. Drawing on work that engages with feminist theory and cultural studies, we suggest that critical reflection and moreover mediation is necessary in order to address burgeoning concerns which have been raised within the NIME discourse in relation to methodological approaches, `diversity and inclusion', `accessibility', and the fostering of rigorous interdisciplinary research.
\end{abstract}





\section{Introduction}
As the International Conference on New Interfaces for Musical Expression (NIME) appears for its twentieth incarnation, we wish to take a moment to draw attention to a variety of interrelated themes that have reached a crisis point within the conference discourse. We see these as exigencies requiring not only critical reflection, but also urgent mediation given that many of the prevailing concerns have been voiced by NIME participants and those from its adjacent communities for many years. Specifically, we will discuss the relationships between prevailing sociopolitical and epistemological struggles within the field through the `planar' framework set out by Born. Born's model interrogates the ways in which multiple temporalities and social planes are at play within historical analyses of musical institutions and their surrounding discourses, and how cultural eras are shaped by these \cite{born2010relational}. While our tone may appear less than optimistic, we believe that in drawing awareness to these structural forces in order to demonstrate how ``powerful discourses or metaphors come to structure musical experiences... conditioning the future musical expressions" \cite[p14]{born2005musical}, there will emerge bountiful opportunities to enrich and transform the field.

\subsection{Declaration of Affiliation}
In this work, we draw analyses not only from what has been documented in the NIME proceedings and the adjacent literature, but also through our lived experiences as participants in the conference. We have been involved with the NIME field in a variety of roles since 2011 and 2010 respectively. We have participated over the years as authors, co-authors, performers, workshop participants, reviewers, and as part of the Program Committee. Between us, we have attended NIME eleven times, and presented our work fourteen times as a combination of papers, posters, and performances. We have both contributed to the first NIME reader \cite{jensenius2017nime}, a collection of works intended to represent a chronological anthology of the field's activities. We have attended and spoken out at NIME town halls, taught NIME-related courses within universities in North and Latin America, and the UK, and we view certain aspects of the NIME discourse as being foundational in our research and practice. 

Yet, we must acknowledge from the outset that our research agendas do not happily align with the dominant narratives that have been reified and reproduced over the course of our engagement with the field. We observe the community emerging from the NIME conference to be extensively heterogeneous in spite of sharing a common interest. This diversity, while having been celebrated since the conference began, does create a problem of a shared perspective and understanding amongst different interest groups \cite{marquez2017}. Overlaps of varying degrees exist between affiliated subdiscplines, conference communities, and academic fields. As such, we refer to `the NIME community' and `NIME research' in the broadest possible senses throughout, allowing for the inclusion of those who may have been faced with barriers to attend or participate in the conference itself. We also hesitantly will use the acronym to refer to both the conference proper, along with its associated research and practice but wish to stress that NIME may suffer from the same existential issues as those applied to `music technology' as a field of study in Boehm's paper on ``[t]he discipline that never was" \cite[p7]{boehm2007discipline}. We are also reluctant to encourage the normalisation of NIME as a noun, often observed to be used interchangeably with digital musical instrument (DMI). These formulations both centre and simultaneously exclude particular themes, notably the embodied and the social aspects of musical activity in the latter case. 

\section{Tracing the N, I, M, and the E}
\begin{quote}The historical present in electronic music and sound cultures is full of contradiction \cite[p79]{rodgers2015cultivating}.\end{quote}

While the very words of New Interfaces for Musical Expression have been contested and explicated individually\footnote{See a case for `ecologies' over `expression' in \cite{gurevich2007expression}.}---and which we ourselves appropriate for the title of this work---it is the I of `interfaces' that Gurevich suggests is often invoked to cement NIME as a field that is historically rooted in the musical inventions from around the turn of the twentieth century \cite{gurevich2016diversity}. This ``audio-technical" \cite[p3]{rodgers2010synthesizing} framing centres the transduction of environmental changes into electrical, and then acoustic signals; the design of---and subsequently engagement with---technical systems or artefacts forms the basis of enquiry. Gurevich also traces an alternative history that is woven through the practices and theoretical writings of what appears to be predominantly male\footnote{Daphne Oram in also included in this short list.} Euro-American composers and experimentalists. He suggests that this framing is more aligned with what would now comprise the methodologies of creative practice research (CPR). While technical reports from the International Computer Music Conference (ICMC) predate the first NIME workshop, explicit formal scientific writing about such new interfaces is generally seen to have emerged concurrently with the conference.  

Arguably, the conference's remit has expanded greatly from a focus on interactive interfaces and novel methods for controlling audio digital signal processing (DSP). The growing number of topics within the conference's Call for Participation evidence this apparent increase, with the eight listed in NIME 2001 growing to twenty-six in 2020\footnote{See \url{https://www.nime.org/past-nimes/}}. While the first few years of the conference saw a fairly continuous growth in participation, the stabilisation in conference attendee numbers that has followed has not been accompanied by epistemological clarity within the research \cite{gurevich2016diversity}. Observations from within the community itself bemoan NIME as an ``artifact-centered field of research" \cite[p1]{elblaus2014nime}, where the discourse has ``tended to focus on implementation issues---with some coverage of performance---but with very scant coverage of wider issues of practice" \cite[p78]{green2016situation}. As Green notes, even where researchers have acknowledged the importance of musical performance practice as a site for evaluation\footnote{This comment refers to an emergent trend within NIME discourse to address the need to evaluate new instruments within a formal framework. This assumes that this kind of evaluation is a primary goal of the research.}, these studies tend to centre on technical objects rather than wider questions of musicality \cite{green2014nime}.  

Concurrently, we see an emergence of calls to the community to address the wider sociopolitical concerns associated with the field. In the title of her paper focusing on questions of representation and authorship, Xamb{\'o} asks ``Who Are the Women Authors in NIME?" \cite{xambo2018women}. Yet \textit{fifteen} years earlier, drawing on feminist theory and science and technology studies (STS), Essl's summons to the NIME community provided an emphatic critique of a discourse which was almost exclusively ``gender unaware, or uncritical" \cite[p26]{essl2003gender}. 

In 2018, the NIME Steering Committee announced a new conference Diversity Statement\footnote{\url{https://www.nime.org/diversity/} \label{refnote}} and Code of Conduct\footnote{\url{https://www.nime.org/code-of-conduct/}}. The former states that one of the goals of NIME is to foster an inclusive environment which invites ``participation from people of all ethnicities, genders, ages, abilities, religions, and sexual orientations"\textsuperscript{\ref{refnote}}. This follows moves within the Association for Computing Machinery's (ACM) Special Interest Group on Computer–Human Interaction (SIGCHI)\footnote{\url{https://sigchi.org/conferences/organizer-resources/organising-a-sigchi-sponsored-} {\url{or-co-sponsored-conference/}}} which encourage that such statements be posted on conference websites, a practice which started appearing in the SIGCHI academic conferences around 2016. Yet, even as NIME 2020 foregrounds the theme of `accessibility', as Skuse notes in her work grounded in critical disability studies, true participation for disabled\footnote{Skuse invokes the social model of disability: the processes of being actively disabled by societal structures, or lack of infrastructures, over the commonly assumed medical model.} musicians can look very different from approaches to accessible technology design which often attempt to ``fix a problem perceived by the non-disabled community" \cite[p4]{skuse}. This point also flags up issues of participation more broadly, and the perpetuated dichotomy between `researchers'---inventors, engineers, designers, musicians, and so on---on one hand, and `users'---imagined recipients of the technical objects---on the other.

\subsection{Socialities, Temporalities, and NIME}
While space prevents us from addressing the various issues that we raise here in substantial depth, we instead offer an overview and synthesis of a now significantly substantial discourse. Certainly, we are conflating a host of issues that have been brought to the NIME community as distinct topics. Yet all of these may be addressed by examining the broader academic, sociopolitical, and historical currents that bind them together. In what follows we ask why, when the ``discipline should be overcoming its initial growing pains" \cite[p80]{gurevich2016diversity}, it seems to have failed to do so. While Gurevich's question was specifically directed towards epistemological and methodological issues within NIME, we extend it here to the wider sociopolitical realms. 

Elblaus and his co-authors have asked, ``what activities are peripheral or external to the domain of NIME development, and is this demarcation made explicit by the researchers or is it implicit?" \cite[p1]{elblaus2014nime}. They suggest that if such boundaries have arisen fortuitously, it is important to interrogate the underlying objectives and motivations of NIME research. In stating our affiliation with NIME, we are positioning ourselves as participants within this discourse. Yet, as demonstrated in Born's framework for understanding the multiple temporalities that are at play within the historical constructs of cultural institutions \cite{born1995rationalizing}, the ``[a]ssess-\\ment of the degree of invention immanent in a particular cultural object... cannot be read off the protagonists' discourse" \cite[p96]{born2010social}. As such, we draw upon not only the official conference proceedings, but also from adjacent publications in related journals, various panels, and associated conferences. Furthermore, we underscore that the institutional situatedness of these platforms should not be the limiting case of NIME's scope and community. Importantly, we attempt to situate this discussion beyond the ``micro-socialities" \cite[p234]{born2010relational} of these domains, attending to a wider cultural and historical analysis. 

Born's extensive project maps out musicological research as an interdisciplinary field, encompassing four topic areas: sociality, temporality, technology, and ontology \cite{born2010relational}. Promoting a relational model, Born's work aims to move beyond the subdisciplinary boundaries that have persisted within musical scholarship by challenging the assumed music/social opposition. As Green has noted, this framework can be applied to help develop new ways of addressing the complexities of the subdiscplinary areas within the context of NIME research \cite{green2014nime}. Specifically, this approach can help to trouble the commitment to binary understandings which separate the ways in which both music and social processes are mediated within NIME research and practice. Regarding sociality, for example, there has been extensive NIME research on the micro-socialities of the first plane---that of musical activity, such as performing with an instrument---and on the second, with the various imagined public groups that engage in these activities, `laptop musicians'\footnote{Both authors perform with laptops but do not associate with this term, nor its related imagined community.} being one such example. On the third plane, only a few clusters of activity within NIME have broached larger issues such as gender, as mentioned previously\footnote{Here we refer to not only the ensuing conversations regarding representation, but moreover the complex ways in which audio-technical discourses become gendered through various mechanisms including language (see \cite{rodgers2010synthesizing}).}. Yet discussions of, for example, race are glaringly absent. On the fourth plane, Born appeals to the social theories of Foucault, Bourdieu, and others, in order to analyse ``the political economy, institutional structures and globalized circulation of music" \cite[p232]{born2010relational}. We suggest that it is now critical to examine how the institutional structures that support NIME research help to sustain and reproduce limiting dominant narratives, and where they might instead be transformational. 

A similar planar scaling is elucidated with respect to the notion of temporalities\footnote{It is beyond the scope of this paper to address the applications of the two other topics of technology and ontology, but it is no doubt obvious that the former should be highly fruitful for NIME scholars to explore.}. This encompasses moving from the highly studied temporal orderings of musical phenomena such as tempo, rhythm, and so on, on the first plane, through musical events (second), genres (third), and finally to the ways in which music and time are co-mediated within various shifting technological, cultural, and sociopolitical conditions, thereby shaping how larger musical epochs evolve \cite{born2010relational}. Significantly, by repeating, privileging, and ultimately reifying certain aesthetic and methodological modalities over the course of its history, there is a danger that the musics and musical activities associated with NIME might fall into the same ``anti-inventive" trappings that Born has identified within other musical institutions, effecting ``a mobile stasis, a capacity to prolong the governing aesthetic by resisting or repressing significant musical change" \cite[p241]{born2010relational}. 

\subsection{Research Diversity in Context}
NIME has always appeared to champion the importance of the sociocultural implications of the numerous innovative and technological developments that it produces. While listed in last place as a relevant topic for the first NIME workshop in 2001, it is evident that the organizers were clearly sympathetic to the idea of addressing the ``artistic, cultural, and social impact of new musical interfaces"\footnote{See \url{https://www.nime.org/2001/} \label{refnote2}.}. Yet how far has NIME research successfully been able to mediate the four social planes set out by Born in order to fully understand the potential reach of its own labours? 

It is uncontroversial to state that sociocultural contexts contribute in shaping research directives, theory, and methods. For example, within the broader human-computer interaction (HCI) domain, how and why such changes take place has been observed and studied. Deemed `waves' \cite{harrison2007three}, these paradigm shifts outline the main ontological and epistemological concerns diachronically, beginning with classical cognitivism, and leading to the current phenomenologically-oriented `third wave' HCI paradigm. These orientations reflect the emphasis given to human experiences and the meaning arising from interaction, rather than mere work or the efficient usage of particular systems. The underlying theoretical basis for such research indeed reflects the emergence of the then recent theories of 4E\footnote{Embodied, embedded, enactive, and extended cognition.} cognition, and the importance given to them by practitioners (see \cite{dourish2004action} for a broad discussion, and \cite{hayes2019beyond} for the musical implications of such a framework). 

It is clear that the main epistemological shift here is from that of the observational to the experiential, and more importantly, towards to the collapse of the epistemic binary between subjectivity and objectivity. Furthermore, within the third HCI paradigm, various perspectives based on feminist approaches \cite{bardzell2010feminist}, queer theory \cite{light2011hci}, and postcolonial studies \cite{irani2010postcolonial} have emerged which embrace plurality, participation, activism, situatedness, and embodiment \cite{harrison2011making}. Practitioners working in these areas have developed critical and potentially transformational approaches to the study of interaction design by troubling its very ontological and epistemological foundations, through what can be referred to as mode of ``agonistic-antagonistic" \cite[p13]{barry2013interdisciplinarity} interdisciplinarity.

Some of this work addresses the problematic issue of ``users as subjects" \cite[p1306]{bardzell2010feminist} by adopting methods such as participatory design. This attempts to deal with the prevalent issue of ``I-methodologies" \cite[p3]{born2016gender} where designers consider themselves to be `typical' users of the objects and artefacts that they design, hence heavily biasing these processes. As NIME is a heavily male dominated community \cite{born2016gender, xambo2018women} gender would be an obvious bias. Aside from study participants, it is also the material conditions---tied to economic and governmental factors---of the working environments and labs, as well as the geographical locations of the researchers themselves that can act as barriers to participation. In CHI, for example, the majority of participating research centres are situated in the United States and Europe. While there are significant communities of HCI practitioners and scholars emerging within Latin America\footnote{\url{https://www.laihc.org/}}, Africa\footnote{\url{https://dl.acm.org/conference/africhi}}, and South Asia\footnote{\url{https://sigchi.org/get-involved/local-chapters/}}, analyses of the citation practices within HCI research has demonstrated how few citations these communities receive (see \cite{henry200720, bidwell2016decolonizing}).

\section{Axes for Action}
In this section, we highlight and knit together some of the more salient issues that have been put forward. This is far from exhaustive but should provide several points of departure for future NIME research.

\subsection{Methodologies}
Examining NIME's origins as a workshop that emerged from the ACM Conference on Human Factors in Computing Systems (CHI) can help to account for the types of methodologies that have been championed since its inception. CHI has been held each year since 1982\footnote{\url{https://dl.acm.org/conference/chi}}, most often within the United States. Members of SIGCHI---who have sponsored the conference since its second year---work in ``fields as diverse as user interface design, human factors, computer science, psychology, engineering, graphics and industrial design, entertainment, and telecommunications"\footnote{\url{https://sigchi.org/about/about-sigchi/}}. Not included in this list of fields are the arts and humanities whose methodologies look very different to those of design, computer science (CS), and engineering. In particular, we note the historic prevalence of the demo format within the CHI conference which has endured through NIME, and its strong links to the ``Demo or Die" \cite{elish2010responsible} maxim adopted by the MIT Media Lab. The NIME community is not unaware of the limitations of the demo format (see \cite{jensenius2016trends}), how it relates to scales of temporalities---moving from promoting `wonder' and `awe' in the `lab' to enduring on the `stage', for example---and how it raises aesthetic and social questions.  

Artistic practice as a mode of academic research has strong links to the the historical centering of ways of knowing and producing bodies of knowledge through practical activity rather than qualitative or quantitative methods. This might include performance, or working with materials \cite{smith2009introduction}, but also the opportunity to imagine novel ways of collective musical activity\footnote{See the LLEAPP project \cite{green2016situation, hayes2019beyond} as an example.}. While NIME 2001 lists ``performance experience reports on live performance and composition using novel controllers"\textsuperscript{\ref{refnote2}} as a key workshop topic, this would not satisfy the type of rigorous methodological approaches associated with the co-constituting and reflexive relationships between theory and practice within CPR\footnote{We are using this as an umbrella term. See \cite{dogantan2016artistic} for a discussion of the nuanced variations between research that is practice-based, practice-led, and so on.} today. CPR has been growing as a recognized approach in certain parts of the world for many decades, although it is important to note that its prevalence and acceptance is extremely varied geographically\footnote{For example, in the United States, the 2019 Alliance for the Arts in Research Universities conference theme was ``knowledges: artistic practice as method", reflecting the ongoing validation of CPR in that country.}. We also acknowledge the messy entanglement between the drive towards the professionalisation of arts practice in the UK as co-emerging with the prevalence of new CPR doctoral degrees.  

The importance of CPR in NIME has been stressed by many, although its potential to transform the field is still largely under-explored and its methods are still underutilised \cite{green2016situation}, and potentially misunderstood. For example, Dahl and his co-authors suggest that ``[p]ractice-based research in new interfaces for musical expression is rooted in the practices of design and performance" \cite[p77]{dahl2016designing}, but seem to circumscribe its application, once again limiting NIME's scope to that of technical objects. Born's framework can act as a catalyst if conceiving what this could entail beyond such artefacts---albeit the design of and performance with---is difficult. Decades of exemplary writing by Oliveros is full of such imagined possibilities for establishing new relationships between sounds and people, artefacts and bodies \cite{oliveros2010sounding}. It is interesting to observe how far from the `computer' of HCI, CHI---traditionally a milieu explicitly \textit{not} engaged with the creative arts and humanities---has managed to advance. Furthermore, as exemplified by Marquez-Borbon and Stapleton \cite{marquez2017}, this type of incoherence of vision also comes to the fore when attempting to study and develop NIME pedagogy, as well as establish performance practices with interactive music systems in general. This is not to suggest that pluralisms are not healthy, but that these should be recognised and deeply engaged with by the community itself.

\subsection{Interdisciplinarity}
It is no surprise that epistemological and methodological issues have surfaced within NIME given that it has always been an interdiscplinary field. Bringing together HCI, CS, and music, NIME can comfortably be situated within the ever growing popular discourse of art-science research \cite{barry2013interdisciplinarity}. Although arguably now scaffolded by decades of research into both HCI and computer music, this resistance to disciplinary siloing should not be assumed to produce identical configurations across the board. Barry and Born outline three potentially intersecting logics of interdiscplinarity that can be helpful \cite{barry2013interdisciplinarity}. While the new and novel `N' of NIME---supported by the prevalence of demos---could be thought of as aligning with their logic of `innovation'---often assumed to be tied to the agendas of funding bodies and governmental agencies, where research is guided and driven specifically by economic demands---there may be cases where this produces both inventive as well as anti-inventive outcomes. 

Similarly, while the `accountability' logic often portrays the arts as in service to the scientific research, particularly in order to make it public facing through communicative means, NIME has often taken a radically reverse approach, bracketing off certain more popular forms of music making from its own discourse. Even more fruitful may be questions of ontology, where, in the third logic, the very question of what constitutes NIME research can be challenged. In this case, we might move away from the problem solving nature of design issues, including the current issue of `evaluation'. Rather, following some of the more radical work appearing within CHI, we might use NIME to problematise existing assumptions that lie within the relationships between sound, people, and technology, specifically \textit{because} NIME research cannot be reduced into the siloed domains of musical, scientific, nor social studies; the unique difference to its precursor conference being, of course, that NIME researchers are also musicians often with deeply developed performance practices. The more inclusive this community allows itself to become, the more these practices can span Born's social planes, uncontestably enriching the field.

\subsection{Quantification}
As a result of methodological struggles within NIME, quantitative methods have prevailed---perhaps due to the dominance of engineering labs and their ability to churn out grant proposals, and hence publish prolifically---despite the variety of other approaches and perspectives at play. This has led to the emergence of various prevalent strands of enquiry including how to evaluate new instruments, whether through user study, or through performance; or how to assess the impact of such instruments through a measure of their longevity. As we have already asserted, these `challenges' may obscure other types of fruitful labour. While NIME hosts its total proceedings online\footnote{\url{https://www.nime.org/archives/}}, it does not archive what may be a much larger number of practice based works---performances, installation, workshops---which have been crucial in the formation of this body of research. There has been some attempt to rectify this with a formal database of musical proceedings from 2019 alone\footnote{ \url{https://github.com/NIME-conference/NIME-bibliography}}. Given that without readily accessible means to reference this work, a huge part of NIME's knowledge base will be excluded from readily being configured within both its history and future formations. Importantly, this exclusion will favour labs and researchers with institutional funding---those who can produce research papers and pay to attend conferences---over independent practitioners.

Through the neoliberal lens of the quantification of the value and impact of research itself, NIME exists within an academic climate where counting matters: for promotion, for tenure, and for gaining future funding. Yet the favouring of particular modes of research, both through methodology as well as citation practices, contributes to shaping both the historical narratives and future directions of the field. As Ahmed has carefully articulated, ``[w]hen citational practices become habits, bricks form walls" \cite[p148]{ahmed2016living}.  Related work on citation practices within CHI demonstrates the often throwaway nature of providing references, where no critical engagement with the work is given beyond its inclusion as mere supporting evidence or validation \cite{marshall2017throwaway}. We should also here consider NIME's discourse surrounding the democratisation of DMIs---a theme which comes up repeatedly---against the financial costs of participating in academic conferences, particularly for students and faculty who are not funded by research labs, and moreover artists who are not supported by academic institutions.

\subsection{`Diversity' and `Inclusion'} 
Resulting from sociocultural changes born out of civil movements addressing race, gender, and economic inequality, Diversity and Inclusion statements have been put in place as ways to establish fair and `safe' spaces for those working and studying within institutions. Given such developments, academic conferences followed suit. However, in spite of these well-meaning approaches, we must continue to take a critical stance and ask whether or how the NIME Diversity Statement provides a productive framework for genuinely increasing the diversity of its community and research outputs. That is, how is the community actively supporting minority or marginalized groups to not only participate, but also to educate and help shape the field? Openness to diversity does not automatically result in it. What types of approaches will the NIME community undertake in its diversity work to avoid not only leaving out those not within positions of power and whose practice lies outwith dominant NIME research \cite{marquez2017}, but also in working towards changing the ways in which such narratives are formed and structure the field itself? In particular, as Rodgers has elucidated in her work on the history of the synthesizer, the effect of language and the construction of narratives can be crucial here: ``audio-technical language and representation, which typically stands as neutral, in fact privileges the perspective of an archetypal Western, white, and male subject" \cite[v]{rodgers2010synthesizing}.

Within the shaping of such narratives, we may attend to the processes of citation practices---as discussed previously---choices of not only keynotes speakers and performers, but also the processes that determine who sits on the Steering Committee and Program Committees of the conference. Furthermore, while there is a `Fair Play' clause in the ethical guidelines for NIME publications\footnote{\url{https://www.nime.org/publication-ethics/}}, which states that papers should be valued in terms of their intellectual content regardless of the authors' gender, belief system, sexual orientation, or race, we are tempted to wonder how biases are mitigated within the make up of the reviewer pool itself. How can we rigorously assess performance work which might be unfamiliar or diverge from the `expected' aesthetics that have become favoured, or those works which suffer from poor documentation due to a lack of resources?\footnote{We have both reviewed well documented performances which failed to deliver in the conference, and vice versa. See our earlier comments on the misleading nature of demos.} How can we continue to support performance work from practitioners who may be the only ones who play their instruments, but simultaneously request anonymous video submissions for double-blind peer review?

Hosting the NIME 2019 conference in Porto Alegre, Brazil is important given that this was the first event to be realised in the entire region of Latin America. This geographical change invites and facilitates regional parties to introduce their work into the larger NIME community, whose work otherwise may be absent when the conference is hosted elsewhere. While this is certainly encouraging, we observe that the participation of Latin American researchers at the 2019 conference turned out to be rather limited. Out of the 88 publications presented at the conference, 18 publications included Latin American authors or co-authors. Thematic lines of research were mainly concerned with the development of new systems and their implementation in a range of diverse usage scenarios. Only a single paper presented a case outside this mainstream, developing a history of Latin American NIME-related activity \cite{lerner2019latin}. This paper is particularly important within discussions of NIME diversity as it illuminates a rich history of instrument development, parallel to the main historical narratives presented within audio-discourse in general. As Lerner states, ``Latin America has configured a fertile ground for experimenting in the NIME field'' \cite[p232]{lerner2019latin} in its recent history, where such developments are not only influenced by their US or European counterparts, but also reflect the local context in which this work takes place. Local technological developments are not mere imitations, but are original and `hybrid' inventions that stand on their own. 

\section{Conclusion}
We have highlighted several ongoing themes and discussions that have come to prominence within the NIME community. While some of the recent NIME discourse appears to be progressive with regards to increasing participation in the field, we stress the importance of a critical approach at all levels. For example, linking together our discussions of quantification with that of gender, we might consider the limitations of the use of metrics which we now see being applied to this domain within NIME discourse (see \cite{xambo2018women}). As Thompson has noted, this process can actually reproduce gender as a rigid category and at the same time prioritise approaches which ultimately turn out to be about ``surface over structure" \cite[p13]{marie}. We wish to challenge the NIME community to respond affirmatively to these ideas, even where the incentive for doing so may be counter-intuitive given the amount of care and labour that will be required, and the increasing commodification of academic research. We echo Green's sentiments in relation to CPR and NIME that ``collective understanding of the research endeavour, rather than a competitive one where researchers themselves are commodities" \cite[p79]{green2016situation} will address both the epistemological and sociocultural fractures that NIME is experiencing.

\section{Acknowledgments}
We thank Owen Green for his suggestions on this text.

%
\bibliographystyle{abbrv}

	\bibliography{nime-references}

\begin{thebibliography}{10}

\bibitem{ahmed2016living}
S.~Ahmed.
\newblock {\em Living a feminist life}.
\newblock Duke University Press, 2016.

\bibitem{bardzell2010feminist}
S.~Bardzell.
\newblock Feminist {HCI}: taking stock and outlining an agenda for design.
\newblock In {\em Proceedings of the {SIGCHI} conference on human factors in
  computing systems}, pages 1301--1310, 2010.

\bibitem{barry2013interdisciplinarity}
A.~Barry and G.~Born.
\newblock Interdisciplinarity: Reconfi gurations of the social and natural
  sciences.
\newblock In {\em Interdisciplinarity}, pages 17--72. Routledge, 2013.

\bibitem{bidwell2016decolonizing}
N.~J. Bidwell.
\newblock Decolonising {HCI} and interaction design discourse: Some
  considerations in planning afri{CHI}.
\newblock {\em XRDS}, 22(4):22–27, June 2016.

\bibitem{boehm2007discipline}
C.~Boehm.
\newblock The discipline that never was: Current developments in music
  technology in higher education in britain.
\newblock {\em Journal of Music, Technology \& Education}, 1(1):7--21, 2007.

\bibitem{born1995rationalizing}
G.~Born.
\newblock {\em Rationalizing culture: IRCAM, Boulez, and the
  institutionalization of the musical avant-garde}.
\newblock University of California Press, 1995.

\bibitem{born2005musical}
G.~Born.
\newblock On musical mediation: Ontology, technology and creativity.
\newblock {\em Twentieth-century music}, 2(1):7--36, 2005.

\bibitem{born2010relational}
G.~Born.
\newblock For a relational musicology: music and interdisciplinarity, beyond
  the practice turn: the 2007 dent medal address.
\newblock {\em Journal of the Royal Musical Association}, 135(2):205--243,
  2010.

\bibitem{born2010social}
G.~Born.
\newblock The social and the aesthetic: For a post-bourdieuian theory of
  cultural production.
\newblock {\em Cultural Sociology}, 4(2):171--208, 2010.

\bibitem{born2016gender}
G.~Born and K.~Devine.
\newblock Gender, creativity and education in digital musics and sound art,
  2016.

\bibitem{dahl2016designing}
L.~Dahl.
\newblock Designing new musical interfaces as research: What’s the problem?
\newblock {\em Leonardo}, 49(1):76--77, 2016.

\bibitem{dogantan2016artistic}
M.~Do{\u{g}}antan-Dack.
\newblock Introduction.
\newblock In M.~Do{\u{g}}antan-Dack, editor, {\em Artistic practice as research
  in music: Theory, criticism, practice}. Routledge, 2016.

\bibitem{dourish2004action}
P.~Dourish.
\newblock {\em Where the action is: the foundations of embodied interaction}.
\newblock MIT press, 2004.

\bibitem{elblaus2014nime}
L.~Elblaus, K.~F. Hansen, and R.~Bresin.
\newblock {NIME} design and contemporary music practice: Benefits and
  challenges.
\newblock In {\em Workshop on Practice-Based Research in New Interfaces for
  Musical Expression, {NIME} 2014}, 2014.

\bibitem{elish2010responsible}
M.~C. Elish.
\newblock Responsible storytelling: communicating research in video demos.
\newblock In {\em Proceedings of the fifth international conference on
  Tangible, embedded, and embodied interaction}, pages 25--28, 2010.

\bibitem{essl2003gender}
G.~Essl.
\newblock On gender in new music interface technology.
\newblock {\em Organised Sound}, 8(1):19--30, 2003.

\bibitem{green2014nime}
O.~Green.
\newblock {NIME}, musicality and practice-led methods.
\newblock In {\em Proceedings of the International Conference on New Interfaces
  for Musical Expression}, 2014.

\bibitem{green2016situation}
O.~Green.
\newblock The situation of practice-led research around {NIME}, and two
  methodological suggestions for improved communication.
\newblock {\em Leonardo}, 49(1):78--79, 2016.

\bibitem{gurevich2016diversity}
M.~Gurevich.
\newblock Diversity in {NIME} research practices.
\newblock {\em Leonardo}, 49(1):80--81, 2016.

\bibitem{gurevich2007expression}
M.~Gurevich and J.~Trevi{\~n}o.
\newblock Expression and its discontents: toward an ecology of musical
  creation.
\newblock In {\em Proceedings of the International Conference on New Interfaces
  for Musical Expression}, pages 106--111, 2007.

\bibitem{harrison2011making}
S.~Harrison, P.~Sengers, and D.~Tatar.
\newblock Making epistemological trouble: Third-paradigm {HCI} as successor
  science.
\newblock {\em Interacting with Computers}, 23(5):385--392, 2011.

\bibitem{harrison2007three}
S.~Harrison, D.~Tatar, and P.~Sengers.
\newblock The three paradigms of {HCI}.
\newblock In {\em Alt. Chi. Session at the {SIGCHI} Conference on human factors
  in computing systems San Jose, California, USA}, pages 1--18, 2007.

\bibitem{hayes2019beyond}
L.~Hayes.
\newblock Beyond skill acquisition: Improvisation, interdisciplinarity, and
  enactive music cognition.
\newblock {\em Contemporary Music Review}, 38(5):446--462, 2019.

\bibitem{henry200720}
N.~Henry, H.~Goodell, N.~Elmqvist, and J.-D. Fekete.
\newblock 20 years of four {HCI} conferences: A visual exploration.
\newblock {\em International Journal of Human-Computer Interaction},
  23(3):239--285, 2007.

\bibitem{irani2010postcolonial}
L.~Irani, J.~Vertesi, P.~Dourish, K.~Philip, and R.~E. Grinter.
\newblock Postcolonial computing: a lens on design and development.
\newblock In {\em Proceedings of the {SIGCHI} conference on human factors in
  computing systems}, pages 1311--1320, 2010.

\bibitem{jensenius2016trends}
A.~R. Jensenius and M.~J. Lyons.
\newblock Trends at {NIME}--reflections on editing “{A} {NIME} reader”.
\newblock {\em In Proceedings of the International Conference on New Interfaces
  for Musical Expression}, page 439–443, 2016.

\bibitem{jensenius2017nime}
A.~R. Jensenius and M.~J. Lyons.
\newblock {\em {A} {NIME} Reader: Fifteen Years of New Interfaces for Musical
  Expression}, volume~3.
\newblock Springer, 2017.

\bibitem{lerner2019latin}
M.~M. Lerner.
\newblock Latin american {NIME}s: Electronic musical instruments and
  experimental sound devices in the twentieth century.
\newblock {\em Proceedings of the International Conference on New Interfaces
  for Musical Expression}, 2019.

\bibitem{light2011hci}
A.~Light.
\newblock {HCI} as heterodoxy: Technologies of identity and the queering of
  interaction with computers.
\newblock {\em Interacting with Computers}, 23(5):430--438, 2011.

\bibitem{marquez2017}
A.~Marquez-Borbon and P.~Stapleton.
\newblock 2015: Fourteen years of {NIME}: The value and meaning of
  ‘community’ in interactive music research.
\newblock In {\em {A} {NIME} Reader}, pages 465--481. Springer, 2017.

\bibitem{marshall2017throwaway}
J.~Marshall, C.~Linehan, J.~Spence, and S.~Rennick~Egglestone.
\newblock Throwaway citation of prior work creates risk of bad {HCI} research.
\newblock In {\em Proceedings of the 2017 {CHI} Conference Extended Abstracts
  on Human Factors in Computing Systems}, pages 827--836, 2017.

\bibitem{oliveros2010sounding}
P.~Oliveros.
\newblock {\em Sounding the margins: collected writings 1992-2009}.
\newblock Deep Listening Publications, Kingston, 2010.

\bibitem{rodgers2010synthesizing}
T.~Rodgers.
\newblock {\em Synthesizing sound: Metaphor in audio-technical discourse and
  synthesis history}.
\newblock PhD thesis, McGill University, 2010.

\bibitem{rodgers2015cultivating}
T.~Rodgers.
\newblock Cultivating activist lives in sound.
\newblock {\em Leonardo Music Journal}, 25:79--83, 2015.

\bibitem{skuse}
A.~Skuse.
\newblock Disabled approaches to live coding, cripping the code.
\newblock {\em Proceedings of the International Conference on Live Coding},
  2020.

\bibitem{smith2009introduction}
H.~Smith and R.~T. Dean.
\newblock Introduction: practice-led research, research-led practice-towards
  the iterative cyclic web.
\newblock {\em Practice-led research, research-led practice in the creative
  arts}, pages 1--38, 2009.

\bibitem{marie}
M.~Thompson.
\newblock Sonic feminisms: Doing gender in neoliberal times.
\newblock In M.~C. Michael~Bull, editor, {\em The Bloomsbury Handbook of Sonic
  Methodologies}. Bloomsbury, forthcoming.

\bibitem{xambo2018women}
A.~Xamb{\'o}.
\newblock Who are the women authors in {NIME}?—improving gender balance in
  {NIME} research.
\newblock In {\em Proceedings of the International Conference on New Interfaces
  for Musical Expression}, pages 174--177, 2018.

\end{thebibliography}

\end{document}